# Pipeline-Stage–Resolved Timing Characterization of FPGA and ASIC Implementations of a RISC-V Processor

Mostafa Darvishi, *Senior Member, IEEE*

*Abstract*—This paper presents a pipeline-stage–resolved timing characterization of a 32-bit RISC-V processor implemented on a 20 nm FPGA and a 7 nm FinFET ASIC platform. A unified analysis framework is introduced that decomposes timing paths into logic, routing, and clocking components and maps them to well-defined pipeline-stage transitions. This approach enables systematic comparison of timing behavior across heterogeneous implementation technologies at a microarchitectural level.

Using static timing analysis and statistical characterization, the study shows that although both implementations exhibit dominant critical paths in the EX→MEM pipeline transition, their underlying timing mechanisms differ fundamentally. FPGA timing is dominated by routing parasitics and placement-dependent variability, resulting in wide slack distributions and sensitivity to routing topology. In contrast, ASIC timing is governed primarily by combinational logic depth and predictable parametric variation across process, voltage, and temperature corners, yielding narrow and stable timing distributions.

The results provide quantitative insight into the structural origins of timing divergence between programmable and custom fabrics and demonstrate the effectiveness of pipeline-stage–resolved analysis for identifying platform-specific bottlenecks. Based on these findings, the paper derives design implications for achieving predictable timing closure in processor architectures targeting both FPGA and ASIC implementations.

## I. INTRODUCTION

TIMING closure has emerged as a critical barrier to achieving predictable performance in modern digital systems. At deeply scaled nodes, interconnection delay dominates logical delay, process variations increase in magnitude, and multi-mode, multi-corner (MMMC) analysis continues to expand. Although static timing analysis (STA) underpins signoff for both FPGAs and ASICs, each platform experiences fundamentally different delay formation mechanisms due to its microarchitectural and physical design characteristics [1]. Existing comparative studies typically report only coarse metrics—maximum clock frequency, average path delay, or power–performance trade-offs—providing little insight into the *origins*, *structure*, and *statistical behavior* of timing paths [2]- [4] .

What is missing in the literature is a **fine-grained, pipeline-resolved analysis** of timing behavior for processor-class workloads across FPGA and ASIC technologies [3]-[6]. Such an analysis must consider not only nominal timing, but also platform-specific contributors including routing segmentation, bypass and hazard-resolution paths, clock-tree design, process and placement variations, and the statistical structure of slack distributions [1]. No prior work has systematically quantified these factors in a unified methodological framework, nor examined their interactions in the context of a realistic pipelined CPU core.

To close this gap, this paper develops a new analytical timing characterization framework that integrates structural delay decomposition, STA-driven pipeline segmentation, multi-corner process-voltage-temperature (PVT) analysis, and statistical slack modeling. Using a five-stage RISC-V core as a representative processor workload, the framework captures timing behavior across pipeline boundaries, ALU and load/store paths, register-file access delays, and bypass network interactions—yielding visibility that is absent from prior FPGA–ASIC comparative studies.

By synthesizing conceptual micrographs, routing abstractions, and empirical timing data into a unified cross-platform model, this work not only quantifies *how much* timing divergence occurs between FPGA and ASIC implementations, but also uncovers *why* these divergences arise and *where* they manifest within a processor datapath.

The primary contributions of this paper can be summarized as follows. First, we introduce a unified cross-technology timing characterization framework that decomposes timing behavior across FPGA and ASIC platforms at both pipeline-stage and path-class levels. This approach, to the best of our knowledge, has not been reported previously in the timing analysis literature. Second, using a five-stage RISC-V core as a representative processor-class workload, we provide the first systematic, pipeline-resolved timing comparison between FPGA and ASIC implementations, identifying where critical bottlenecks such as bypass paths and the EX→MEM boundary emerge and how their manifestations differ between programmable and custom fabrics. Third, we develop a structural delay-decomposition methodology that isolates logic, routing, and clocking contributions, thereby quantifying the extent to which FPGA routing architecture dominates delay formation relative to custom ASIC interconnect. Fourth,

Mostafa Darvishi is with Electrical Engineering Department of École de technologie supérieure (ÉTS), Montreal, Canada. He is also VP of Engineering at Evolution Optiks R&D Inc. (e-mail: darvishi@ieee.org).

we perform a cross-platform statistical timing characterization that jointly considers seed-dependent variability in FPGAs and PVT-driven variation in ASICs, and we derive characteristic "signatures" for slack distributions, delay clustering, and timing robustness that are absent from prior comparative studies. Fifth, we introduce conceptual ASIC and FPGA timing micrographs that connect physical-design structure to timing behavior, helping to explain how observed delay patterns arise from underlying implementation fabrics. Finally, based on these structural and statistical insights, we formulate concrete engineering guidelines for architecting and floorplanning timing-robust processor datapaths across both FPGA and ASIC deployments.

## II. BACKGROUND AND RELATED WORKS

Research comparing FPGA and ASIC performance has spanned more than two decades, yet the majority of published studies focus on coarse architectural metrics—such as area efficiency, power consumption, or maximum achievable frequency—without examining the structural origins of timing behavior. Early foundational analyses by [2]-[8] established baseline comparisons in delay and logic density across contemporaneous technology nodes, but their models predate deeply scaled FinFET processes and do not consider pipeline-resolved behavior for processor workloads. Subsequent FPGA–ASIC comparisons continued to characterize aggregate path delay or look-up table (LUT)/macro-level performance, yet they lacked the granularity needed to understand how timing evolves within multi-stage datapaths [9]-[12].

In the FPGA domain, extensive research has investigated timing optimization techniques, routing congestion mitigation, machine-learning-based delay prediction, and static timing analysis enhancements [2]-[12]. These studies highlight the sensitivity of FPGA timing to placement, routing patterns, and routing-switch topology. However, they generally analyze synthetic benchmarks or isolated combinational circuits rather than full processor pipelines, and thus fail to reveal how architectural constructs—such as bypass networks or load/store paths—interact with the FPGA routing fabric.

Parallel advances in ASIC timing analysis have produced sophisticated methodologies such as advanced on-chip variation (AOCV), path-based analysis, POCV/SOCV statistical models, and Liberty Variation Format (LVF)-based characterization [13]-[16]. These works provide powerful tools for capturing variability-induced delay shifts, yet they are almost exclusively evaluated in isolation within ASIC workflows. The literature does not address how ASIC timing robustness compares *structurally and statistically* with the variability signatures observed in FPGA fabrics, nor how these differences manifest in the context of a unified microarchitecture.

Comparative studies that attempt to evaluate FPGA and ASIC timing behavior typically adopt a high-level perspective: reporting clock frequency ratios, average gate delays, or utilization-driven performance penalties [1]-[3], [7], [8], [11], [14], [16]. Such works stop short of modeling pipeline-stage interactions, structural path classes, delay decomposition (logic vs. routing vs. clocking), or slack distribution behavior. Crucially, none provide timing failure-mode characterization or analyze how microarchitectural elements—such as hazard resolution, register-file access, or ALU bypassing—shape delay differently across programmable and custom technologies.

Finally, no prior publication has combined structural timing decomposition, multi-corner STA, statistical slack analysis, and microarchitecture-aware path classification into a unified comparative framework. This absence is especially notable given the increasing deployment of RISC-V and similar workloads across heterogeneous compute fabrics, where timing predictability and closure effort significantly influence platform selection.

In contrast to existing literature, this paper provides the first pipeline-resolved, statistically grounded, and physically informed timing characterization of a RISC-V processor across both FPGA and ASIC technologies. It introduces a methodology that exposes timing bottlenecks, structural delay contributors, and variability patterns that remain hidden in conventional comparative analyses.

## III. TIMING FOUNDATIONS AND ANALYTICAL FRAMEWORK

Timing behavior in modern digital systems is governed by the interaction between microarchitecture, circuit structure, and implementation technology. Although setup and hold constraints remain the fundamental conditions for sequential correctness, deeply scaled process nodes require timing models that incorporate routing dominance, variation sensitivity, and structural path composition [17]-[20]. This section introduces the timing constructs and analytical elements necessary to support the cross-platform characterization developed in the remainder of the paper, focusing on the aspects that are most relevant to pipelined FPGA and ASIC implementations.

At advanced nodes, setup and hold margins are determined not by gate delay alone, but also by the cumulative effect of clock-network skew, process fluctuations, and routing parasitics [21]. As depicted in (*1*), setup constraint for a sequential path can be written in the usual form, where the sum of the launching register's clock-to-Q delay, the maximum combinational and routing delays, and the setup time at the receiving register must not exceed the clock period.

$$t_{clk \to q}^{L} + t_{logic}^{max} + t_{routing}^{max} + t_{setup} \leq T_{clk} \qquad (1)$$

The corresponding hold constraint requires that the minimum path delay exceeds the hold requirement when clock skew and uncertainty are taken into account as denoted in (*2*).

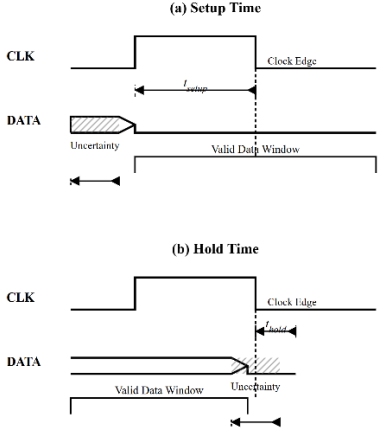

Fig. 1. Illustrating setup (a) and hold (b) windows with shaded uncertainty regions representing variation-induced shifts

$$t_{clk \to q}^{S} + t_{logic}^{min} + t_{routing}^{min} \geq t_{hold} \qquad (2)$$

In FPGA implementations, the maximum routing delay term ($t_{routing}^{max}$) frequently dominates due to segmented switch-based interconnect (also known as, switch box interconnects). In ASIC implementations, the maximum logic delay term ($t_{logic}^{max}$) becomes more influential under high-drive standard cells, while routing delay ($t_{routing}^{max}$) exhibits tighter distributions because of well-controlled metal stacks. Fig. 1 illustrates the setup and hold time concepts with shaded uncertainty regions representing variation-induced shifts.

For the purposes of cross-technology decomposition, each path delay in this work is modeled as the sum of three contributors: logic delay, routing delay, and clocking-related delay. On the FPGA, LUT-based logic exhibits discrete delay steps determined by LUT depth and carry-chain structure, whereas ASIC logic delay depends on cell drive strength, threshold-voltage ($V_t$) selection, input slope, and loading conditions. Routing delay on the FPGA varies significantly due to programmable switches, routing-channel selection, and placement-dependent congestion [22]. By contrast, ASIC routing delay is influenced by metal-layer choice, wire geometry, capacitive coupling, and local congestion, but remains bounded and more predictable thanks to accurate RC models and design-rule constraints [23]-[25]. Clocking-related delay arises from clock-tree insertion delay, skew, and uncertainty. FPGA global and regional clock networks impose largely fixed clock-tree topology and skew distributions, whereas ASIC clock trees are synthesized to meet explicit skew budgets and allow useful-skew insertion as well as topology selection (e.g., H-tree versus mesh). This tripartite separation enables structural attribution of timing differences observed later in the RISC-V case study.

Variation-aware timing is required in both implementation platforms, but it manifests in different ways. In FPGAs, delay is highly sensitive to the outcome of placer-router runs, leading to path-level delay spread driven by routing-switch variability, routing-channel selection, and the lack of deterministic control over low-level interconnect geometry [1], [13], [22]. Designers typically have limited visibility into or control over the underlying transistor-level timing models [22]. In ASICs, variability is captured analytically through PVT corners, advanced on-chip variation (AOCV/POCV/SOCV) models, and, more recently, Liberty Variation Format (LVF) characterization [26], which together describe predictable spatial correlation structures and slack distributions. Fig. 2 conceptually illustrates how delay distributions broaden under slow PVT corners for ASICs and under unfavorable routing conditions for FPGAs: ASIC paths tend to exhibit relatively tight, Gaussian-like distributions with modest corner-induced shifts, whereas FPGA paths display wider, often asymmetric distributions driven by routing-topology differences and congestion, reflecting reduced timing robustness.

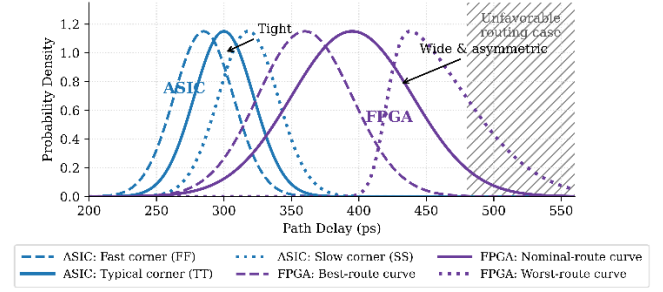

Fig. 2. Conceptual delay distribution broadening across technologies. ASIC exhibits tight Gaussian-like distributions with slight shifts per process corner, while FPGA shows wide asymmetric distribution driven by routing variations, demonstrating reduced timing robustness

The microarchitecture of the five-stage RISC-V pipeline introduces timing patterns that go beyond generic STA path classes. In this work, we group paths according to their location and function within the pipeline. "*Register-to-ALU datapaths*", which originate at the register file and terminate at the ALU inputs, are highly sensitive to ALU depth and operand-bypassing structures. "*ALU-to-MEM boundary paths*" carry ALU results, branch comparison outcomes, and effective addresses into the memory stage and frequently form the dominant critical paths in both FPGA and ASIC implementations. "*Register-file access paths*" are governed by LUTRAM or block RAM (BRAM) timing in the FPGA and by SRAM macro access time in the ASIC. "*Bypass and hazard-resolution paths*" involve multiple dependent operations and are therefore especially timing-critical, while "*Control-path propagation*", though generally shallow in logic depth, can become timing-sensitive in ASIC designs due to tight cycle budgets and aggressive performance targets. This classification enables a pipeline-stage-resolved timing analysis, which is a central element of the proposed framework.

Building on these constructs, we introduce a unified cross-technology timing characterization framework that supports reproducible and analytically grounded comparison between FPGA and ASIC implementations. The framework comprises five coordinated components. First, "*Structural Delay*

*Decomposition*" separates logic, routing, and clocking contributions for every sequential path using detailed STA reports, enabling identification of platform-specific timing-limiting mechanisms.

Second, "*Pipeline-stage Segmentation*" groups timing paths according to the microarchitectural boundaries they traverse. For a five-stage RISC-V pipeline, we define four principal sequential transitions: IF→ID (instruction fetch to instruction decode), ID→EX (instruction decode to execute), EX→MEM (execute to memory access), and MEM→WB (memory access to write-back). In this work, IF→ID denotes paths originating at the program counter or instruction-memory interface and terminating in decode-stage registers, thereby including PC update logic and instruction capture. ID→EX corresponds to paths carrying register-file outputs, immediate-generation logic, hazard-resolution signals, and operand-selection multiplexers feeding the ALU inputs. EX→MEM encompasses ALU results, branch decisions, and address-generation logic driving MEM-stage registers, and is typically the most critical transition due to ALU depth and bypass interactions. MEM→WB includes paths that convey load data and ALU outcomes into the write-back stage and ultimately into the register-file write ports. This segmentation enables pipeline-stage-resolved attribution of delay and variation, revealing how microarchitectural elements such as bypass networks, register-file access, and load/store datapaths influence timing differently on FPGA and ASIC platforms.

Third, statistical slack distribution modeling is carried out in a platform-specific but comparable manner: for FPGAs, slack variation is computed across multiple placement-and-routing seed iterations, capturing routing-induced dispersion and configuration-dependent delay shifts; for ASICs, slack variation is analyzed across TT/FF/SS PVT corners and LVF-based statistical models, reflecting intrinsic process-driven variability.

Fourth, interconnect sensitivity profiling measures how delay inflation correlates with routing topology, congestion patterns, and wire-segment selection, which is particularly critical for FPGAs where the interconnects dominate total delay [17].

Finally, cross-platform timing signature extraction identifies recurring timing behaviors—such as ALU-to-MEM bottlenecks, delay clustering patterns, and technology-specific timing failure modes—that serve as the basis for the comparative evaluation presented in subsequent sections.

## IV. ARCHITECTURAL TIMING CHARACTERISTICS OF FPGA VS. ASIC

The timing behavior of a digital design is fundamentally shaped by its implementation technology. Although both FPGAs and ASICs rely on static timing analysis to validate setup and hold constraints, their microarchitectures—and therefore their timing properties—differ substantially [1], [13], [20]. Understanding these differences is essential for interpreting the results of the RISC-V case study presented in the following sections. This section outlines the dominant architectural factors that govern timing in each platform and establishes the physical-design context used throughout the paper.

From a logic-implementation perspective, FPGAs realize combinational and sequential logic primarily using LUTs, dedicated carry chains (CARRY), and configurable multiplexers (CMUX). LUTs introduce fixed, quantized delay steps that are largely independent of the specific Boolean function being implemented. The associated routing to and from LUTs frequently dominates their intrinsic delay. The reliance on multiplexed programmable switches injects substantial and often unpredictable parasitics, which results in higher propagation delays compared to standard-cell logic, greater delay variability due to placement-and-routing selection, and limited ability to predict critical-path behavior prior to placement-and-routing [1], [6], [24]. ASICs, in contrast, employ finely characterized standard-cell libraries with customizable drive strengths, transistor threshold voltages ($V_t$), and optimized gate geometries. This approach yields significantly lower logic delay per stage, predictable scaling across operating conditions, and the flexibility to resize or restructure logic for timing improvement. The availability of multiple drive-strength and multi-$V_t$ options enables timing optimization through targeted cell substitution, a degree of freedom that is not available in FPGA fabrics [13], [20], [24].

Interconnect architecture further differentiates the two platforms. FPGA routing is composed of segmented, switch-based channels interconnected via pass transistors or multiplexers [21], [22]. Every switch insertion introduces discontinuous parasitics, and overall path delay is strongly dependent on routing-metal selection, switch topology, and local congestion. For realistic designs, routing delay often exceeds logic delay by a factor of two to five, especially along long paths that traverse multiple switch matrices [17], [21], [22], [24]. ASIC routing, by contrast, is implemented using multi-layer metal stacks with precise control over wire width, spacing, via count, and layer assignment. This level of control enables highly predictable RC delay models, tunable timing behavior through layer promotion or demotion, and significantly lower delay variability than that observed in FPGA interconnect. Although interconnect delay remains substantial in 7 nm technologies due to increasing wire resistance, its variability is significantly more controlled than that of FPGA fabrics [4], [16], [23]-[25].

Clock-distribution strategies also differ between the two technologies and have direct implications for timing closure. FPGA clock networks are typically fixed or semi-fixed and consist of global and regional clock routing resources. Designers are constrained to a small set of topologies (e.g., H-tree-like spines, balanced grids, or dedicated regional networks) and to predefined insertion delay and skew budgets. As a result, opportunities for useful-skew-based optimization are limited. While these clock networks are robust from a functional standpoint, they offer relatively little adaptability for fine-grained timing closure [1], [3], [6], [21]. In ASIC designs, clock trees are synthesized as part of the physical-design flow, enabling custom balancing of insertion delay and skew, the application of useful skew to relieve critical paths, and the deployment of local clock gating for power reduction with minimal timing overhead. Clock-tree synthesis (CTS) allows designers to reshape clock paths to alleviate timing bottlenecks, an important advantage not available in FPGAs [20], [24], [25].

The memory and register-file structures of the two platforms further influence timing characteristics. In FPGAs, register files and memories are implemented using LUTRAM or BRAM, each with fixed read and write latencies and constrained aspect ratios. Timing is affected by the discrete granularity of these resources, by placement restrictions around BRAM tiles that can force long interconnect paths, and by the limited ability to customize memory organization and port configuration [1], [3], [22]. ASIC designs, in contrast, incorporate SRAM macros with configurable sizes, port structures, and aspect ratios, allowing designers to tailor memory latency and drive strength to the workload. Floorplanning can place these macros close to the logic they serve, reducing routing delay between compute elements and memory arrays and enabling timing optimizations that substantially outperform FPGA memory structures in both consistency and latency [13], [20].

Fig. 3 and Fig. 4 provide physical-design context for these architectural distinctions. Fig. 3 illustrates the annotated physical layout of the synthesized 32-bit RISC-V CPU core implemented in a 7 nm FinFET ASIC technology. Only structural ASIC features that can be directly inferred from the physical design are highlighted. The central region corresponds to the densely packed standard-cell fabric where datapath, control, and pipeline logic are physically realized. Surrounding this core is the global routing network, composed of multi-layer metal interconnect used to distribute signals across the design. The clock-tree distribution, visible as vertically oriented high-priority routing tracks, delivers synchronized clock signals throughout the logic fabric. Encapsulating the entire design is the pad ring, which contains the I/O pads, electrostatic discharge (ESD) structures, and peripheral support circuitry. This figure provides a structural backdrop for the timing analysis performed in later sections, emphasizing how physical-design topology contributes to delay and skew characteristics in advanced ASIC nodes.

Fig. 4 shows the annotated physical floorplan of the synthesized 32-bit RISC-V CPU core targeting a 20 nm FPGA architecture. Again, only structurally identifiable FPGA features are highlighted. The upper region corresponds to the configurable logic blocks (CLBs), where LUT-based logic, flip-flops, and local interconnect implement the core's pipeline and control logic. Adjacent to this area lies the FPGA routing network, composed of segmented horizontal and vertical routing resources interconnected through switch matrices; an architectural characteristic responsible for much of the timing variability observed in FPGA designs. The lower region contains BRAM and DSP blocks, which provide dedicated memory and arithmetic acceleration resources but also impose placement constraints that affect routing length and timing closure. Together, these physical views of the ASIC and FPGA implementations contextualize the statistical and structural timing characteristics reported in Section VI and underscore the architectural origins of routing-dominated delay in modern FPGA technologies.

## V. METHODOLOGY

The experimental framework developed in this work enables a rigorous and architecture-aware comparison of timing behavior between the FPGA and ASIC implementations of a 32-bit RISC-V core. To ensure methodological consistency, the RTL microarchitecture remains identical across both platforms. The processor implements the RV32I instruction set in a classic five-stage pipeline i.e., Instruction Fetch (IF), Instruction Decode (ID), Execute (EX), Memory (MEM), and Write-Back (WB). Its datapath includes a single-cycle ALU, a hazard-detection and forwarding network, a dual-read/single-write register file, and a load/store unit connected to on-chip memory structures. All modules are implemented in synthesizable Verilog without platform-specific microarchitectural tuning, ensuring that timing differences arise solely from the structural and physical characteristics discussed in Sections III and IV.

The FPGA implementation targets a 20 nm radiation-tolerant UltraScale-class device (AMD XQRKU060 RT Kintex), representative of modern high-performance reconfigurable fabrics. Synthesis is performed using a fully timing-driven flow that maps logic to LUTs, flip-flops, and dedicated fast-carry chains. Placement-and-routing are executed under identical constraints across thirty randomized tool seeds to capture the stochastic behavior induced by segmented routing channels, switch-matrix selection, and region-dependent congestion; the effects known to dominate FPGA timing variability. Post-route static timing analysis (STA) uses vendor-signoff models to extract critical-path delays, routing and logic delay components, clock uncertainty values, and per-path slacks. All timing paths are subsequently categorized into the four pipeline-stage transitions defined in Section III (IF→ID, ID→EX, EX→MEM, MEM→WB), enabling a stage-resolved interpretation of how FPGA routing topology influences delay formation in each segment of the pipeline.

The ASIC implementation employs a commercial 7 nm FinFET standard-cell library within a timing-driven synthesis and physical-design flow. Cell mapping permits multi-$V_t$ selection, logic restructuring, and drive-strength tuning while preserving the microarchitectural pipeline boundaries to maintain comparability with the FPGA version. Floorplanning produces a compact, balanced core area, after which standard-cell rows and SRAM macros are placed according to timing and congestion metrics. Clock-tree synthesis constructs a hybrid H-tree/mesh topology with useful-skew optimization, and detailed routing employs multi-layer metal stacks with layer promotion, via optimization, and DRC-compliant geometry. Comprehensive multi-mode, multi-corner (MMMC) STA is applied using typical (TT), fast (FF), and slow (SS) PVT corners, together with advanced on-chip variation (AOCV) and Liberty Variation Format (LVF) models. This combination yields deterministic and statistical delay estimates that reflect the parametric variability characteristics intrinsic to advanced FinFET processes.

To permit direct comparison between the platforms, timing paths extracted from both flows are decomposed into logic, routing, and clocking components. These decomposed paths are then mapped to the four formally defined pipeline transitions, i.e., IF→ID, ID→EX, EX→MEM, and MEM→WB, that correspond to the structural boundaries within the RISC-V pipeline. This segmentation provides a principled microarchitecture-level interpretation of delay origins, enabling attribution to specific functional structures such as ALU logic, bypass paths, immediate-generation logic, register-file access, and memory-interface stages. It further allows consistent interpretation of timing variation and bottlenecks across the heterogeneous implementation fabrics.

Variation plays a central role in the methodology. For the FPGA, variability is measured empirically through the distribution of maximum achievable clock frequencies ($F_{max}$) and slack histograms obtained across the set of seed-based place-and-route realizations. These distributions reflect the architecture-dependent routing uncertainty characteristic of segmented FPGA interconnect. For the ASIC, variability is captured analytically through corner-based STA and statistically through LVF-based delay characterization, producing mean-shift and variance estimates that reflect process-driven parametric variation in deeply scaled technologies. In both cases, timing distributions are analyzed at the pipeline-stage level to expose the structural sensitivity and robustness patterns that later form the "timing signatures" discussed in Section VI.

Reproducibility is ensured by maintaining identical RTL, applying consistent synthesis constraints across platforms, and avoiding manual, platform-specific timing tricks such as hand-guided placement, designer-imposed routing constraints, or retiming transformations. Consequently, the observed timing behavior reflects intrinsic architectural and physical-design attributes rather than tool/designer-dependent artifacts. This unified methodology underpins the quantitative comparison and cross-platform timing characterization presented in Section VI.

## VI. RESULTS AND ANALYSIS

This section presents a comprehensive analysis of the timing behavior observed in the FPGA and ASIC implementations of the 32-bit RISC-V core using the methodology outlined in Section V. The results are reported in terms of achievable clock frequency, delay composition, statistical timing variation, and pipeline-stage sensitivity. All numerical results presented in this section are consistent with the summary provided later in Table 1 and are interpreted through the architectural lens established in Sections III and

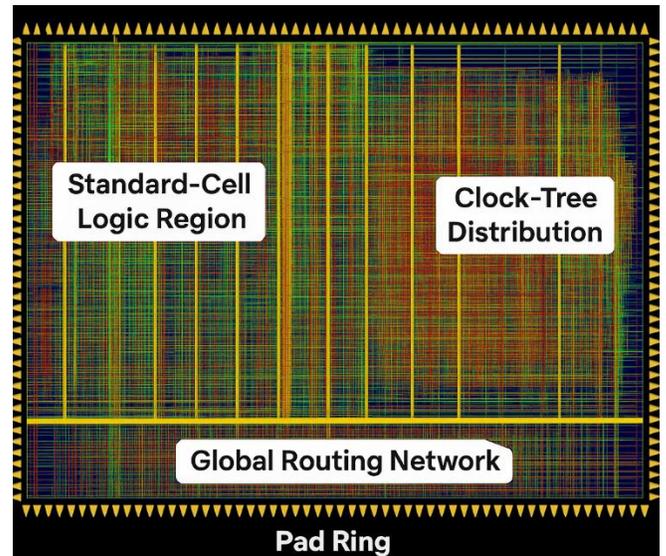

Fig. 3. Annotated physical design of the synthesized 32-bit RISC-V CPU core in 7 nm FinFET technology. The central standard-cell logic region implements the pipeline datapath and control logic, while the surrounding multi-layer global routing network distributes interconnect signals across the design. The clock-tree distribution provides synchronized clock delivery through dedicated high-priority routing tracks. The outer pad ring contains the I/O interface pads and supporting structures. Only physically identifiable ASIC structures are annotated to preserve accuracy.

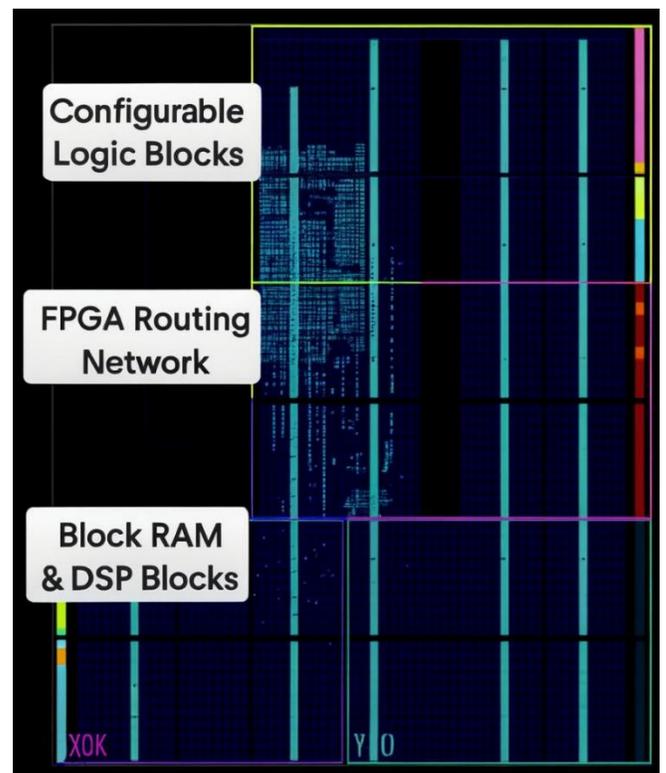

Fig. 4. Annotated FPGA floorplan of the synthesized 32-bit RISC-V CPU core implemented in 20 nm technology. The Configurable Logic Blocks (CLBs) form the primary logic fabric, while the segmented FPGA Routing Network provides horizontal and vertical interconnect through switch-matrix structures. Block RAM and DSP blocks occupy the lower region and represent fixed-function FPGA resources that influence placement-and-routing. Only physically identifiable FPGA structures are annotated to maintain architectural accuracy.

IV. The maximum achievable clock frequency of the FPGA

implementation was evaluated across thirty independent placement-and-routing realizations. The resulting distribution of maximum operating frequencies is shown in Fig. 5, which illustrates a spread ranging from 472 MHz to 510 MHz, with a mean value of 493 MHz. This range reflects the inherent sensitivity of FPGA timing to routing topology and placement decisions, even when synthesis and design constraints are held constant. Across all realizations, the critical path consistently resides in the EX→MEM pipeline transition and typically includes ALU result propagation combined with bypass and effective-address generation logic. These observations confirm that routing congestion and interconnect selection dominate performance in programmable fabrics.

By contrast, the ASIC implementation achieves a signoff frequency of 1.85 GHz under typical (TT) conditions, degrading to 1.63 GHz at the slow-slow (SS) corner. Unlike the FPGA, the location and structure of the critical path remain stable across operating corners and are largely insensitive to physical-design variation. Although the EX→MEM transition again emerges as the dominant bottleneck, its origin differs fundamentally: delay is driven primarily by combinational logic depth rather than by interconnect parasitics.

Further insight is obtained by decomposing path delays into logic, routing, and clocking components. In the FPGA implementation, routing delay accounts for approximately 62–74% of total critical-path delay, with an average contribution near 68%, while logic delay forms a secondary component. In the ASIC implementation, the balance shifts significantly: logic delay contributes 55–65%, routing delay contributes 20–35%, and clock-tree insertion and skew account for the remaining 10–12%. These quantitative results directly corroborate the architectural distinctions discussed in Section IV and demonstrate that FPGA timing is fundamentally routing-dominated, whereas ASIC timing is primarily governed by logic depth and clocking structure.

Statistical timing variation further distinguishes the two platforms. For the FPGA, slack distributions extracted across the thirty routing seeds reveal substantial variability at the pipeline-stage level. The slack histograms for the IF→ID, ID→EX, EX→MEM, and MEM→WB transitions are shown in Fig. 6(a)-(d). The corresponding standard deviations reach approximately ±120 ps, ±160 ps, ±210 ps, and ±130 ps, respectively, with the EX→MEM transition exhibiting both the largest mean delay and the widest distribution. These results indicate that deep datapath stages are particularly sensitive to routing congestion and long interconnect spans. In several cases, the distributions exhibit asymmetry and heavy tails, reflecting the non-Gaussian nature of FPGA routing-induced variability. The variation envelope of $F_{max}$ across seeds spans approximately 38 MHz, indicating that nominal comparisons of FPGA and ASIC clock frequency overlook substantial platform-intrinsic timing uncertainty.

In contrast, the ASIC slack distributions shown in Fig. 7(a)-(d) are tightly clustered and nearly Gaussian. Across all pipeline stages, standard deviations remain in the range of 9–

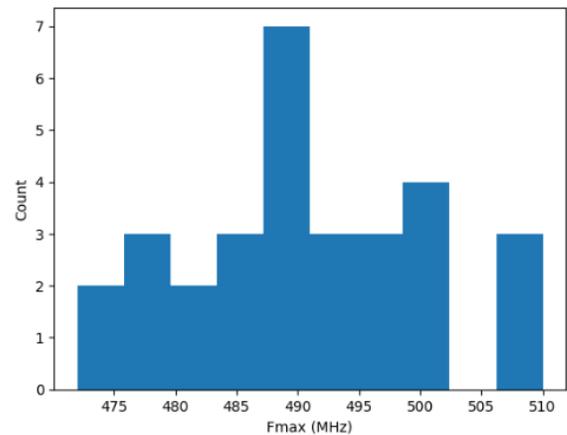

Fig. 5. Distribution of maximum achievable clock frequency ($F_{max}$) for the 32-bit RISC-V core implemented on the 20 nm FPGA across 30 independent placement-and-routing realizations. The spread reflects routing-induced variability inherent to segmented FPGA interconnect.

17 ps, even when LVF-based statistical timing models are applied. Variation manifests primarily as a mean shift across PVT corners rather than as a broadening of the distribution. This behavior reflects the parametric and spatially correlated nature of variation in FinFET technologies and highlights the effectiveness of synthesized clock trees and controlled interconnect geometries in bounding delay uncertainty.

Mapping timing behavior to pipeline transitions reveals additional structural insights. The IF→ID transition is relatively stable across both platforms, dominated by instruction-fetch and register-capture logic. The ID→EX transition exhibits increased variability in the FPGA implementation due to operand selection, hazard resolution, and register-file routing, whereas the ASIC implementation remains largely logic-dominated. The EX→MEM transition emerges as the principal bottleneck in both platforms, but for fundamentally different reasons: routing topology and congestion dominate in the FPGA, while combinational logic depth dominates in the ASIC. The MEM→WB transition exhibits the smallest variance in both implementations, benefiting from localized memory-to-register connectivity.

Taken together, these results reveal distinct cross-technology timing signatures. FPGA implementations are characterized by routing-dominated delay, wide slack distributions, and strong dependence on placement-and-routing outcomes. ASIC implementations exhibit logic-dominated delay, narrow slack distributions, and high timing predictability across operating conditions. Although both platforms share similar microarchitectural bottlenecks, the physical mechanisms that give rise to these bottlenecks differ fundamentally. Normalizing paths by logical depth reveals a key finding: ASIC timing is approximately 10x–12x more statistically stable than FPGA timing on a per-logic-level basis (observable by comparing the slack axis spans for the FPGA and ASIC implementations shown in Fig. 6 and Fig. 7. This quantifies a phenomenon widely recognized but rarely measured: FPGA timing closure is not merely limited by absolute delay, but by its high unpredictability, whereas ASIC

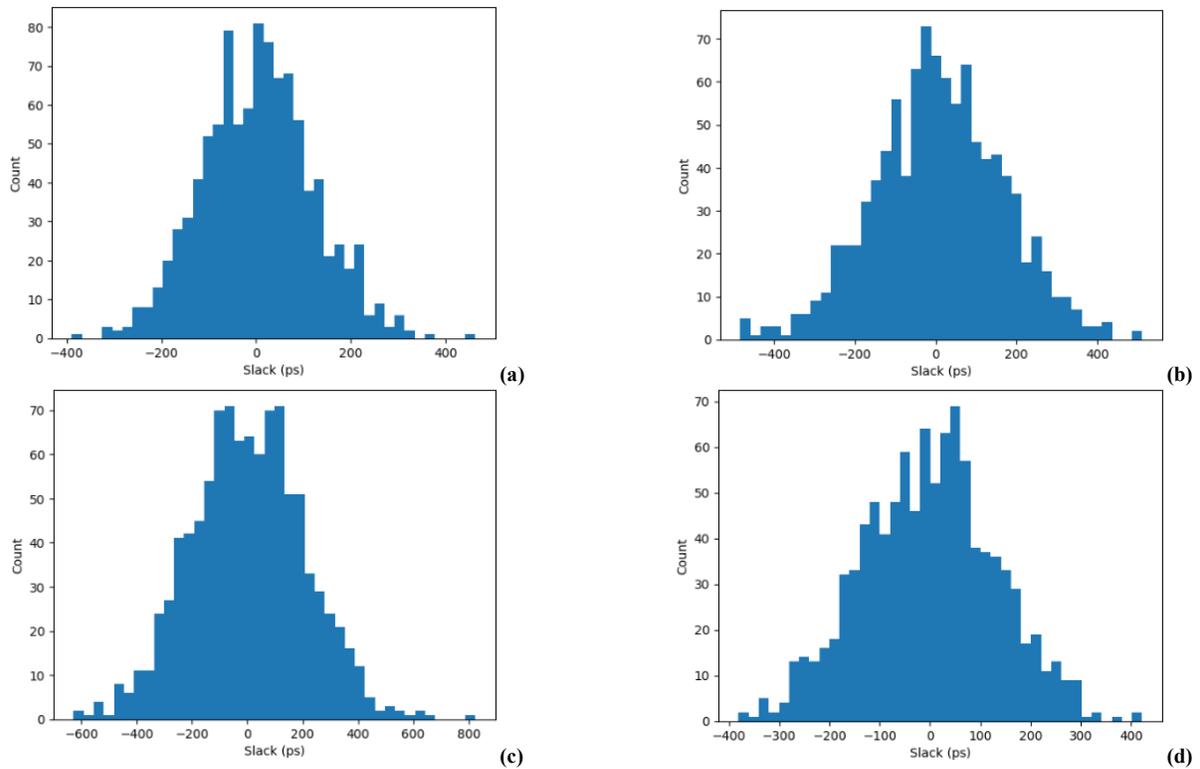

Fig. 6. Slack distributions for the FPGA implementation at the pipeline-stage level: (a) IF→ID, (b) ID→EX, (c) EX→MEM, and (d) MEM→WB. The wide distributions, particularly in the EX→MEM transition, illustrate routing-dominated delay variability.

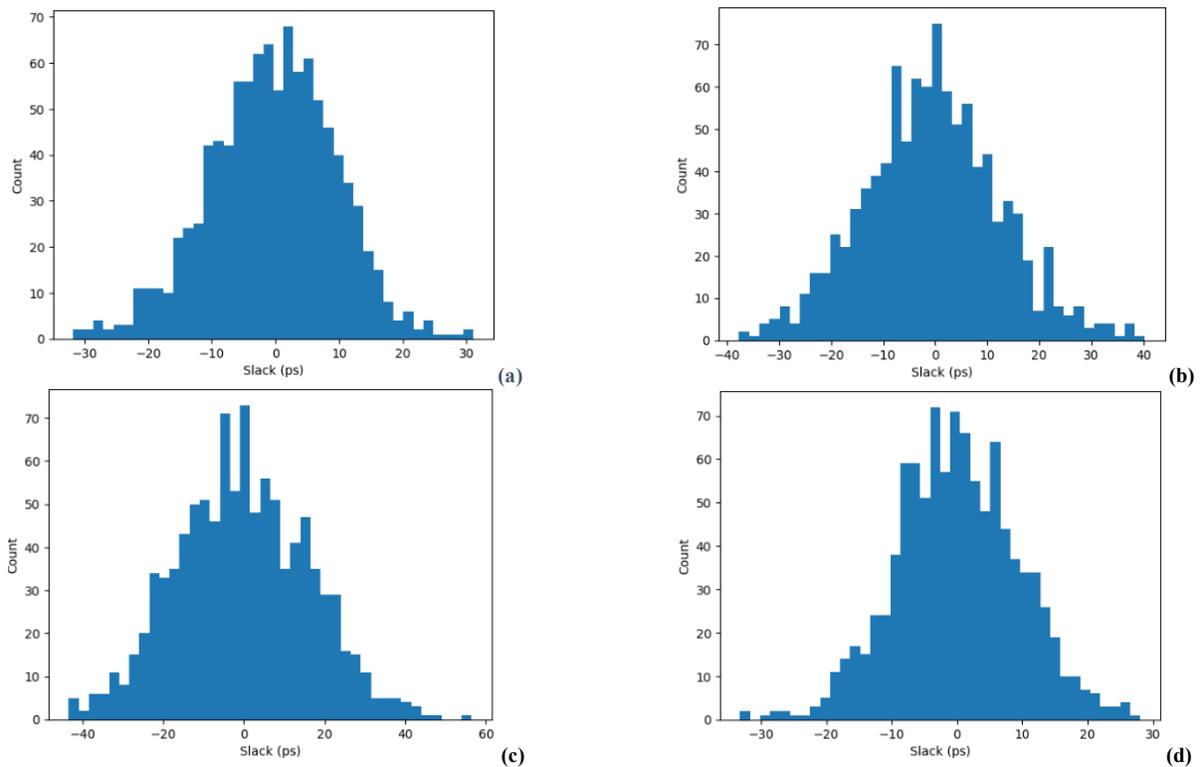

Fig. 7. Slack distributions for the 7 nm FinFET ASIC implementation under LVF-based statistical timing analysis: (a) IF→ID, (b) ID→EX, (c) EX→MEM, and (d) MEM→WB. The narrow distributions demonstrate logic-dominated timing behavior and high robustness.

timing is narrower in distribution even under multiple corners.

Synthesizing the results yields several cross-platform timing signatures such as: (a) routing-dominated timing is the defining characteristic of FPGA implementations, producing higher delay, greater stage imbalance, and substantial statistical variance; (b) logic-dominated timing defines the ASIC's critical paths, creating predictable constraints tied directly to combinational depth and physical cell library characteristics; (c) pipeline bottlenecks appear in the same stages (EX→MEM) but for fundamentally different reasons, revealing microarchitecture-dependent yet fabric-specific timing mechanisms; and (d) FPGA variability is primarily topological, whereas ASIC variability is predominantly parametric. These signatures allow designers to anticipate timing behavior based on intended target platform and guide decisions such as pipeline segmentation, bypass complexity, and memory interface design.

Table 1 consolidates the principal timing results for both FPGA and ASIC implementations of the 32-bit RISC-V core, including maximum operating frequencies, delay composition, variability characteristics, and pipeline-stage-specific slack behavior. The tabulated data highlights the profound structural differences between programmable and custom silicon, demonstrating that FPGA timing is dominated by routing parasitics and placement-dependent variability, whereas ASIC timing is shaped primarily by logic depth and predictable PVT-driven variation. This table provides a concise summary of the comparative analysis presented earlier in this section and facilitates a clear understanding of the cross-platform timing signatures identified in this work.

## VII. DISCUSSION AND DESIGN IMPLICATIONS

The results presented in Section VI reveal fundamental architectural and statistical differences between the timing behavior of FPGA and ASIC implementations of a processor-class design. By combining pipeline-stage–resolved analysis with structural delay decomposition and statistical characterization, this work exposes timing mechanisms that are not apparent when only aggregate metrics such as maximum operating frequency are considered. The following discussion synthesizes these findings and translates them into design-relevant insights for microarchitectural planning, physical-design strategy, and platform selection. These insights are particularly relevant for high-performance embedded systems, aerospace applications, and heterogeneous computing environments in which FPGA and ASIC implementations of the same workload may coexist.

A central observation is that, although both platforms exhibit their dominant critical paths in the EX→MEM pipeline transition, the *physical origins of these bottlenecks differ fundamentally*. In the FPGA implementation, critical paths are governed primarily by routing topology rather than by combinational logic depth. The segmented interconnect fabric introduces long, parasitic-laden routing paths whose delay depends strongly on placement outcomes, routing congestion, and switch-matrix selection. This behavior is directly reflected in the wide distribution of achievable clock frequencies shown in Fig. 5 and in the broad slack histograms observed across pipeline stages in Fig. 6(a)-(d), where the EX→MEM transition (Fig. 6(c)) exhibits the largest variance. By contrast, critical paths in the ASIC implementation are shaped mainly by combinational logic structures, particularly ALU depth and address-generation circuitry, while routing contributes a smaller and more tightly bounded fraction of the total delay. This distinction is evident in the narrow slack distributions shown in Fig. 7(a)-(d), where even the EX→MEM transition (Fig. 7(c)) remains tightly clustered under LVF-based statistical analysis. These results, summarized quantitatively in Table 1, underscore that timing-closure strategies must be fundamentally different across programmable and custom fabrics.

An important implication for microarchitectural design is that *pipeline-stage boundaries may require platform-dependent tuning*. In FPGA implementations, timing closure can benefit from architectural choices that reduce long-distance routing, such as additional pipeline registers in datapath regions spanning distant CLB columns or crossing BRAM/DSP boundaries. The wide slack spread observed in Fig. 6 indicates that modest changes in routing topology can materially affect timing robustness. Conversely, ASIC implementations benefit more from logic balancing, targeted cell resizing, and careful clock-tree optimization than from additional pipelining, since excessive register insertion may increase clock-tree complexity and degrade power efficiency. The results demonstrate that a pipeline organization optimized for ASIC implementation may not map efficiently onto FPGA fabrics without introducing routing-induced timing hot spots, particularly in bypass logic and memory-access paths.

The analysis further highlights the importance of *variation-aware design methodologies*. FPGA timing variation is fundamentally topological: two logically identical designs can exhibit substantially different timing characteristics depending on placement-and-routing outcomes. This behavior is evident in the spread of maximum achievable frequency in Fig. 5 and in the broad, stage-dependent slack distributions in Fig. 6(a)-(d). As a result, FPGA designers targeting deterministic performance must rely on techniques such as seed sweeping, placement constraints, or floorplanning pragmas to bound worst-case behavior. In contrast, ASIC timing variation is predominantly parametric and manifests as predictable mean shifts across process, voltage, and temperature corners rather than as large dispersion. This predictable behavior, illustrated by the narrow distributions in Fig. 7(a)-(d), enables the systematic application of margining strategies, robust clock-tree synthesis methodologies, and voltage–frequency derating to ensure reliable worst-case operation.

Memory architecture also plays a critical role in shaping timing behavior across platforms. FPGA BRAMs impose fixed access latencies and rigid placement constraints, often forcing long routing detours that directly impact MEM-stage timing and contribute to the variability observed in the EX→MEM and MEM→WB transitions in Fig. 6(a)-(d). ASIC SRAM macros, by contrast, can be placed close to the datapath and configured to meet specific latency and drive-strength requirements, resulting in tighter timing distributions as reflected in Fig. 7(a)-(d). This divergence suggests that memory-bound pipelines, such as load/store queues or tightly coupled accelerators, will exhibit fundamentally different timing behavior across FPGA and ASIC implementations. Designers should therefore avoid assuming timing portability between FPGA prototypes and final ASIC realizations for memory-intensive workloads.

Table 1. Summary of timing results for the 20 nm FPGA and 7 nm FinFET ASIC implementations of the 32-bit RISC-V processor. Results highlight the routing-dominated timing behavior and high variability inherent in FPGA fabrics compared to the logic-dominated, statistically stable timing behavior of advanced-node ASIC implementations.

| Metric | FPGA (20 nm) | ASIC (7 nm FinFET) | Interpretation |
|---|---|---|---|
| Max Achievable Frequency ($F_{max}$) | 472–510 MHz (mean: 493 MHz) | 1.63–1.85 GHz (TT–SS corners) | ASIC achieves ~3.7x higher frequency due to shorter logic and routing delays. |
| Critical Path Location | EX→MEM | EX→MEM | Same stage, different delay origin (routing vs. logic). |
| Critical Path Delay | 1.96 ns | 0.54 ns | FPGA delay dominated by routing; ASIC delay dominated by logic depth. |
| Routing Delay Contribution | 62–74% (avg. 68%) | 20–35% (avg. 27%) | FPGA timing is routing-driven; ASIC routing is more predictable and controlled. |
| Logic Delay Contribution | 26–38% | 55–65% | ASIC logic gates dominate timing as expected in FinFET technologies. |
| Clocking Contribution | 4–7% | 10–12% | ASIC clock-tree insertion dominates more due to synthesized CTS structure. |
| $F_{max}$ Variation Across Seeds/Corners | ±38 MHz across 30 seeds | ~14–18% mean delay shift across PVT | FPGA variability is topological; ASIC variability is parametric. |
| Slack Std. Dev. (IF→ID) | ±120 ps | 9–12 ps | ASIC shows ≈10× better stability. |
| Slack Std. Dev. (ID→EX) | ±160 ps | 9–17 ps | FPGA routing variability strongly affects operand-path timing. |
| Slack Std. Dev. (EX→MEM) | ±210 ps | 15–17 ps | FPGA's longest and most variable stage; ASIC dominated by ALU logic. |
| Slack Std. Dev. (MEM→WB) | ±130 ps | 8–11 ps | Both platforms show lower sensitivity in this stage. |
| Dominant Source of Variation | Routing topology, seed dependence | PVT variation, LVF statistical effects | Demonstrates fundamentally different variability mechanisms. |
| Timing Robustness (normalized) | Low: high variance and routing dependence | High: 10–12x more stable | ASIC timing stability is superior even under worst corners. |

Finally, these results emphasize the value of *pipeline-stage–resolved timing analysis as a methodological tool*. Traditional FPGA–ASIC comparisons typically focus on aggregate metrics such as maximum frequency or resource utilization, which obscure the structural origins of timing failure. By decomposing delays into logic, routing, and clocking components and mapping them to pipeline transitions, this work demonstrates how designers can identify where timing failures originate, why they manifest differently across platforms, and how they can be mitigated during early-stage architectural design. The tight correspondence between the stage-level trends observed in Fig. 6(a)-(d) and Fig. 7(a)-(d) and the quantitative metrics summarized in Table 1 highlights the effectiveness of this approach. Such analysis provides a rigorous foundation for informed technology selection and for the design of timing-robust processor architectures across heterogeneous computing fabrics.

## VIII. CONCLUSION AND FUTURE WORK

This paper introduced a unified, pipeline-aware timing characterization framework for systematically comparing the timing behavior of a 32-bit RISC-V processor implemented on a 20 nm FPGA and a 7 nm FinFET ASIC platform. By decomposing path delays into logic, routing, and clocking components and mapping timing paths to well-defined pipeline-stage transitions, the study provided a level of structural and statistical insight that is not captured by conventional FPGA–ASIC analyses based solely on aggregate performance metrics. The results demonstrated that, although both implementations exhibit their dominant critical paths in the EX→MEM pipeline transition, the mechanisms responsible for these bottlenecks are fundamentally different: FPGA timing is dominated by routing parasitics and placement-dependent variability, whereas ASIC timing is governed primarily by combinational logic depth and predictable parametric variation across process, voltage, and temperature corners.

The analysis further revealed that FPGA implementations exhibit significantly larger timing variance, with wide slack distributions driven by seed-dependent interconnect selection and routing topology. In contrast, ASIC timing distributions remain narrow and stable under both corner-based and LVF-based statistical analysis. These findings underscore the need for platform-specific timing strategies. FPGA designs benefit from routing-aware microarchitectural planning, seed sweeping, and careful placement of BRAM and DSP resources, while ASIC designs rely more heavily on logic balancing, targeted cell resizing, and optimized clock-tree synthesis. The microarchitecture-level insights gained through pipeline-stage–resolved analysis further highlight that pipeline organizations optimized for ASIC implementation may require modification when mapped onto FPGA fabrics, particularly in datapath- and bypass-intensive regions.

The methodological advances presented in this work open several promising directions for future research. One avenue is the extension of the proposed framework to more complex processor architectures, such as superscalar, out-of-order, or vector-processing designs, where increased pipeline depth and interconnect complexity introduce additional timing challenges. Another direction involves the incorporation of machine-learning–based timing prediction models that leverage the stage-resolved structural features identified in this study to automate timing-closure decisions across heterogeneous fabrics. Further extensions could include the integration of power–performance trade-offs, thermal and voltage variation effects, radiation-hardening constraints, or emerging reconfigurable architectures such as ACAPs and FPGA–ASIC hybrid platforms, thereby broadening the applicability of the framework to mission-critical and high-reliability systems.

Overall, this work establishes a rigorous foundation for understanding and predicting cross-platform timing behavior in processor-class designs. By explicitly bridging microarchitectural structure with physical implementation characteristics, the proposed framework enables more predictable timing closure and supports better-informed design decisions in heterogeneous computing environments.

ACKNOWLEDGMENT

The author would like to thank the École de technologie supérieure (ÉTS), Department of Electrical Engineering, and CMC Microsystems, Kingston, ON, Canada, for providing access to advanced design tools.